\documentclass[epj]{webofc}
\usepackage[varg]{txfonts}   
\usepackage{hyperref}
\newcommand*{\tabref}[1]{Table~\ref{tbl:#1}}
\newcommand*{\tablab}[1]{\label{tbl:#1}}
\renewcommand*{\eqref}[1]{Eq.~(\ref{eq:#1})}

\newcommand*{\figref}[1]{Fig.~(\ref{fig:#1})}
\newcommand*{\figlab}[1]{\label{fig:#1}}

\usepackage[utf8]{inputenc}
\usepackage{siunitx}

\wocname{\includegraphics[width=0.25cm,clip]{logoARENA} ARENA2018}
\wocname{ARENA2018}
\woctitle{ARENA2018}
\woctitle{\includegraphics[width=0.25cm,clip]{logoARENA} ARENA2018}
\begin{document}
\title{Acoustic detection of neutrinos in bedrock}
%
%

\author{\firstname{Wladyslaw Henryk} \lastname{Trzaska}\inst{1}\fnsep\thanks{\email{wladyslaw.h.trzaska@jyu.fi}}
\and \firstname{Kai} \lastname{Loo}\inst{1}
\and \firstname{Timo} \lastname{Enqvist}\inst{1}
\and \firstname{Jari} \lastname{Joutsenvaara}\inst{2}
\and \firstname{Pasi} \lastname{Kuusiniemi}\inst{1}
\and \firstname{Maciej} \lastname{Slupecki}\inst{1}}


\institute{Department of Physics, University of Jyv\"askyl\"a, Finland
\and
           Kerttu Saalasti Institute, University of Oulu, Finland
          }

\abstract{
We propose to utilize bedrock as a medium for acoustic detection of particle showers following interactions of ultra-high energy neutrinos. With the density of rock three-times larger and the speed of sound four-times larger compared to water, the amplitude of the generated bipolar pressure pulse in rock should be larger by an order of magnitude. Our preliminary simulations confirm that prediction. Higher density of rock also guarantees higher interaction rate for neutrinos. A noticeably longer attenuation length in rock reduces signal dissipation. The Pyh\"asalmi mine has a unique infrastructure and rock conditions to test this idea and, if successful, extend it to a full-size experiment.  
}
\maketitle
%

\section{Introduction}
\label{intro}
The pioneering 1957 paper by G.A. Askariyan \cite{1_Askariyan} has outlined a mechanism converting energy deposited by an ionising particle stopping in a liquid into a hydrodynamic pressure-wave. One of the widely investigated applications of this phenomenon is the search for the characteristic bipolar pressure pulses (BIP; see \figref{fig-BIP}) induced by cascades following interaction of ultra-high energy neutrinos (E $\geq$ 10\textsuperscript{18} eV) with water or ice. These two target materials are the only ones currently used or intended for the future deployment of acoustic sensors. At some point salt domes and sub-arctic permafrost were also discussed as a possible detection material, but no large-scale measurements were attempted. For an in-depth review of the field see e.g. \cite{2_Lahmann:2016qmc}. To our knowledge, utilization of bedrock for that purpose has never been considered before. 

\begin{figure}[h]
\centering
\includegraphics[width=6cm,clip]{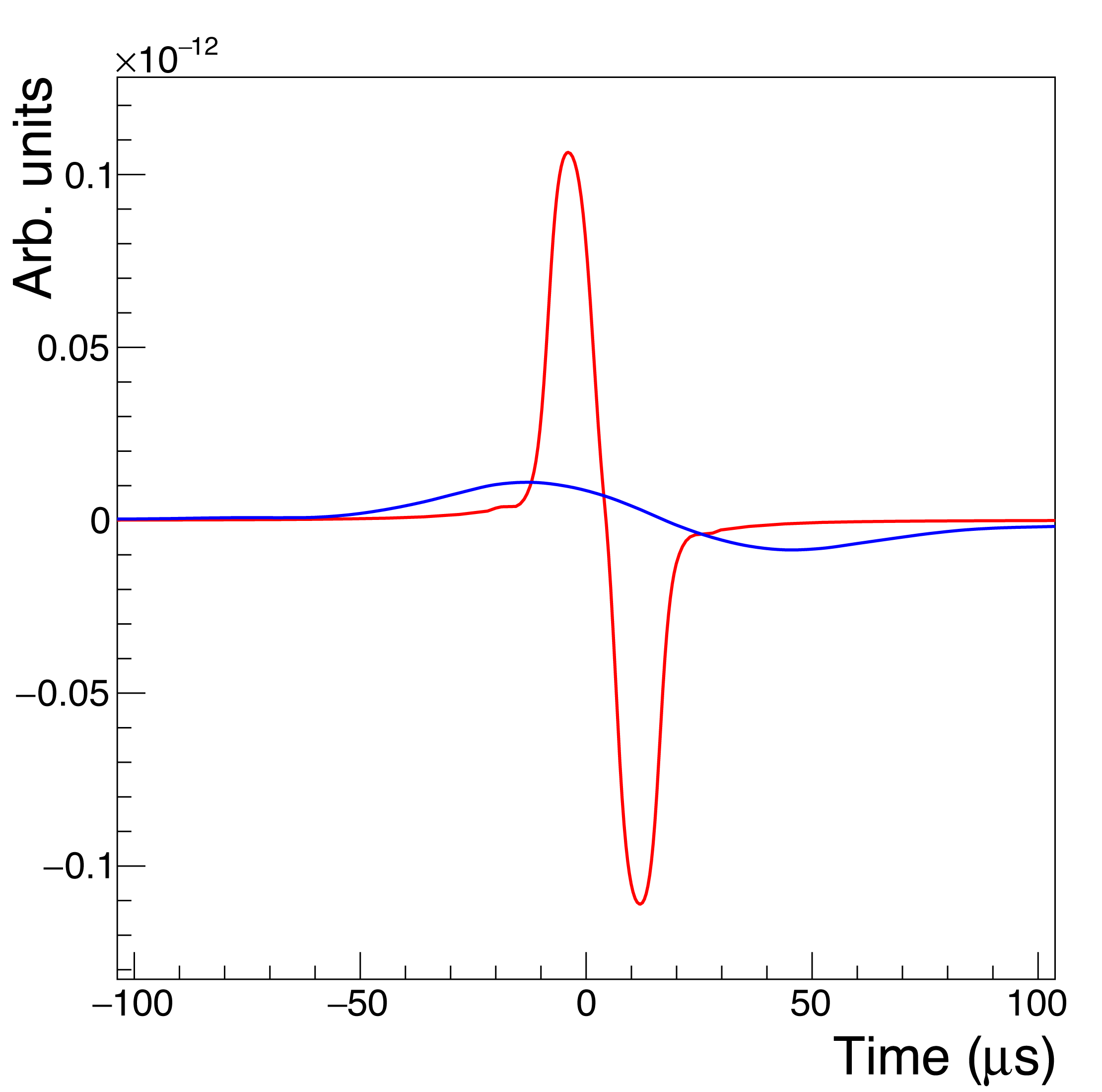}
\caption{Comparison of bipolar pressure pulses generated by the same ultra-high-energy particle-induced shower in water (blue line) and in rock (red line). Please note one order of magnitude difference is signal amplitude and three-fold difference in pulse length/ frequency.}
\figlab{fig-BIP}
\end{figure}

\section{Advantages of rock}
\label{intro}
A dense and ridged material, e.g. granite rock, has a number of advantages for generation and propagation of sound waves produced by interactions of ultra-high energy neutrinos. Since the properties of ice and water are similar, especially in terms of density, for brevity we have restricted our comparison exclusively to hard rock and water. The main reasons favouring rock over water are: a three-times larger density, four-times larger speed of sound, and a longer attenuation length for sound waves. 

The advantages of a denser material are two-fold: neutrino cross section scales with the number of nuclei/electrons per unit volume, and the interaction length of charged particles in a denser medium is shorter. The former means more interactions within the same volume and the latter translates into a stronger signal as the heated mass is more compact and the generated temperatures higher (see \figref{fig-edepsim}). Our simplified, preliminary simulations imply that, compared to water, BIP generated in rock would be stronger by an order of magnitude (\figref{fig-BIP}). This is a convincing argument in favour of further R\&D in that direction. In particular, development of signal generation and propagation models in rock should be accompanied by actual test measurements. 

\begin{figure}[h]
\centering
\includegraphics[width=11cm,clip]{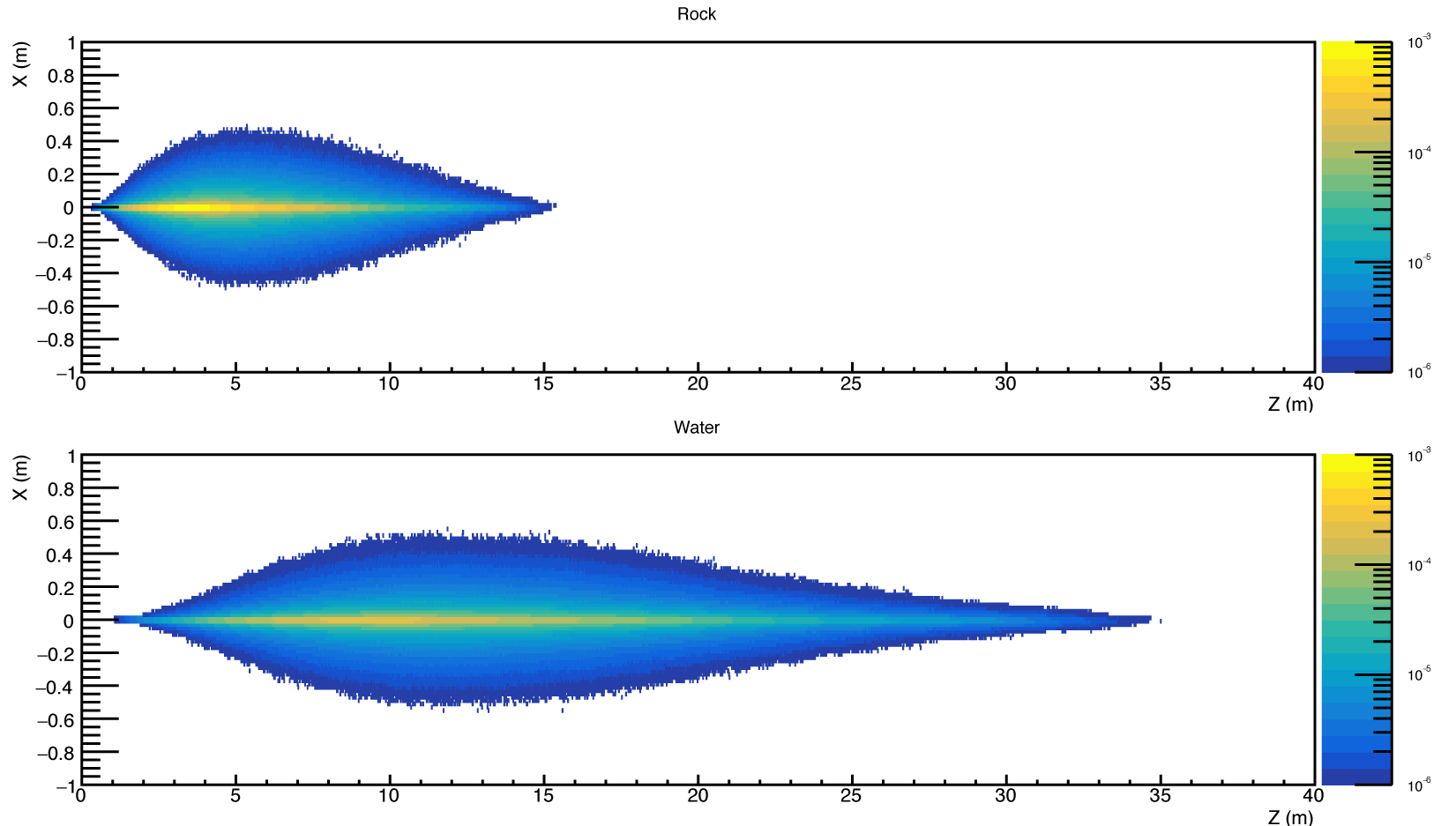}
\caption{Probability distribution of energy deposition by a shower of charged particles generated through an interaction of the same high-energy particle in rock (top) and in water (bottom).}
\figlab{fig-edepsim}
\end{figure}

Some of the reasons why the search for acoustic fingerprints of EeV neutrinos in bedrock has never been proposed could have been the lack of suitable infrastructures and concerns about the lack of uniformity of the geological formations. Also, the potential costs of producing a network of deep deposition holes in granite bedrock, covering an area of several square-kilometres, appears to be beyond practical consideration.

In principle, salt is a form of rock. It has similar, albeit lower, density (80\% of that of rock), speed of sound (77\% of that of rock), etc. Nevertheless, compared to bedrock, salt formations are relatively rare and, with exception of salt domes \cite{14_Saltzberg:2006gv}, deposited in thin layers making them less suited for large-scale neutrino detection. In addition, the acoustic parameters of salt depend heavily on pressure \cite{15_Bonner}.

The main advantage of ice, yet another material similar to rock, is the possibility to conduct acoustic and optical measurements in the same large volume. The main disadvantages are the low density and the need to conduct the experiments at the South Pole – the only place where the suitable ice layer is available.

\section{Infrastructure of the Pyh\"asalmi mine}
\label{infra}
The Pyh\"asalmi mine in Finland is the deepest metal mine in Europe. The main level of the mine, where the crusher, repair and maintenance workshops, storage, and social facilities are located is at the depth of 1400 m. In terms of safety and technology Pyh\"asalmi is one of the best in the world. The ore deposit is located within a compact, cylindrically-shaped vertical volume, surrounded on all sides by a strong, nearly-monolithic rock. The advantages of using the mine for scientific purposes were evaluated, for instance, as part of the LAGUNA and LAGUNA-LBNO Design Studies \cite{3_Laguna}\cite{4_Trzaska:2011zz}\cite{5_Trzaska:2010zz}. 

\begin{figure}[h]
\centering
\includegraphics[width=8cm,clip]{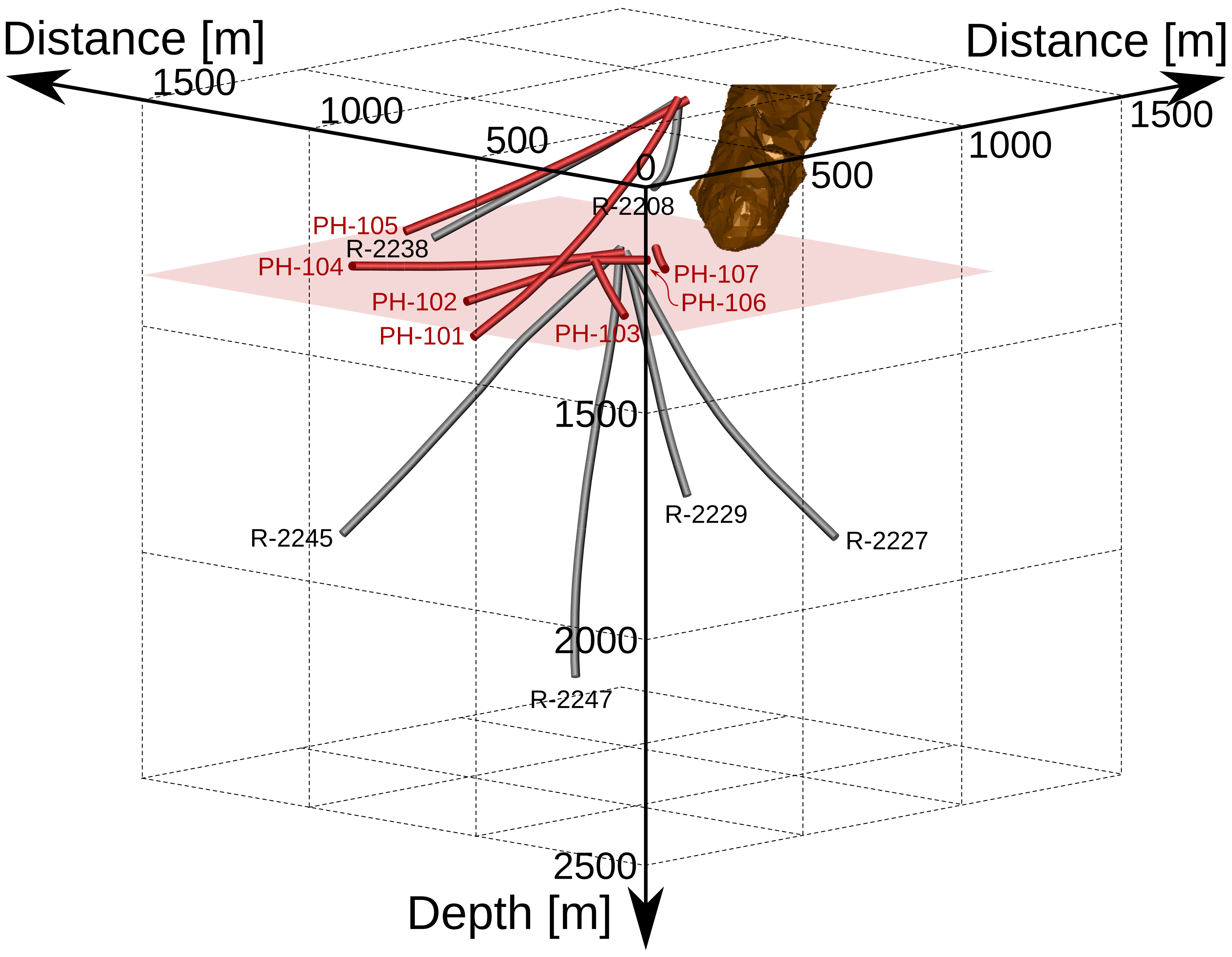}
\caption{The outline of the test boreholes made during the LAGUNA-LBNO Extended Site Investigation. The explored area covers the volume of about 1~km\textsuperscript{3} reaching from the depth of around 1300 m down to 2500 m. The total length of just the PH boreholes (marked in red) is 3.5 km.}\figlab{fig-drillholes}
\end{figure}

Over the years the search for new ore deposits has generated an extensive network of boreholes reaching far out and deep down the surrounding area. These boreholes, with very well documented geological profile, are now available for scientific research. Especially relevant to the acoustic detection were the drillings conducted within the Extended Site Investigations at Pyh\"asalmi aimed at locating the best rock formations for excavation of the giant caverns needed for the LAGUNA-LBNO project \cite{3_Laguna}. As illustrated in \figref{fig-drillholes}, the explored area covers the volume of about 1 km\textsuperscript{3} reaching from the depth of around 1300 m down to 2500 m. The total length of the new boreholes is 3.5 km.

\section{Science in the mine}
\label{science}
Ore deposits in the Pyh\"asalmi mine are expected to last till fall 2019 but the use of the mine infrastructure for other commercial activities is expected to continue. Callio \cite{6_Callio} – the organization promoting and coordinating the transfer to the post-excavation phase – is also providing laboratory base for the ongoing and future scientific projects. 

The main physics activities in the mine are the cosmic-ray experiment EMMA \cite{7_Kuusiniemi:2018vbz}, located at the depth of 75 m in Callio Lab1, and the new low-background laboratory \cite{8_Bezrukov:2018bme} – the Callio Lab2 – at the depth of 1430 m (4100 m.w.e.). The experimental hall of Callio Lab2 (\figref{fig-labphotos}) has the area of about 120 m\textsuperscript{2} and the maximum height of 9 metres. At the entrance to the experimental hall there is a loading dock, sealed at both ends with steel doors. It is large enough to accommodate a full-size truck. The area of Callio Lab2 is well lit and ventilated, equipped with power outlets and fast internet connection. 

\begin{figure}[h]
\centering
\includegraphics[width=13cm,clip]{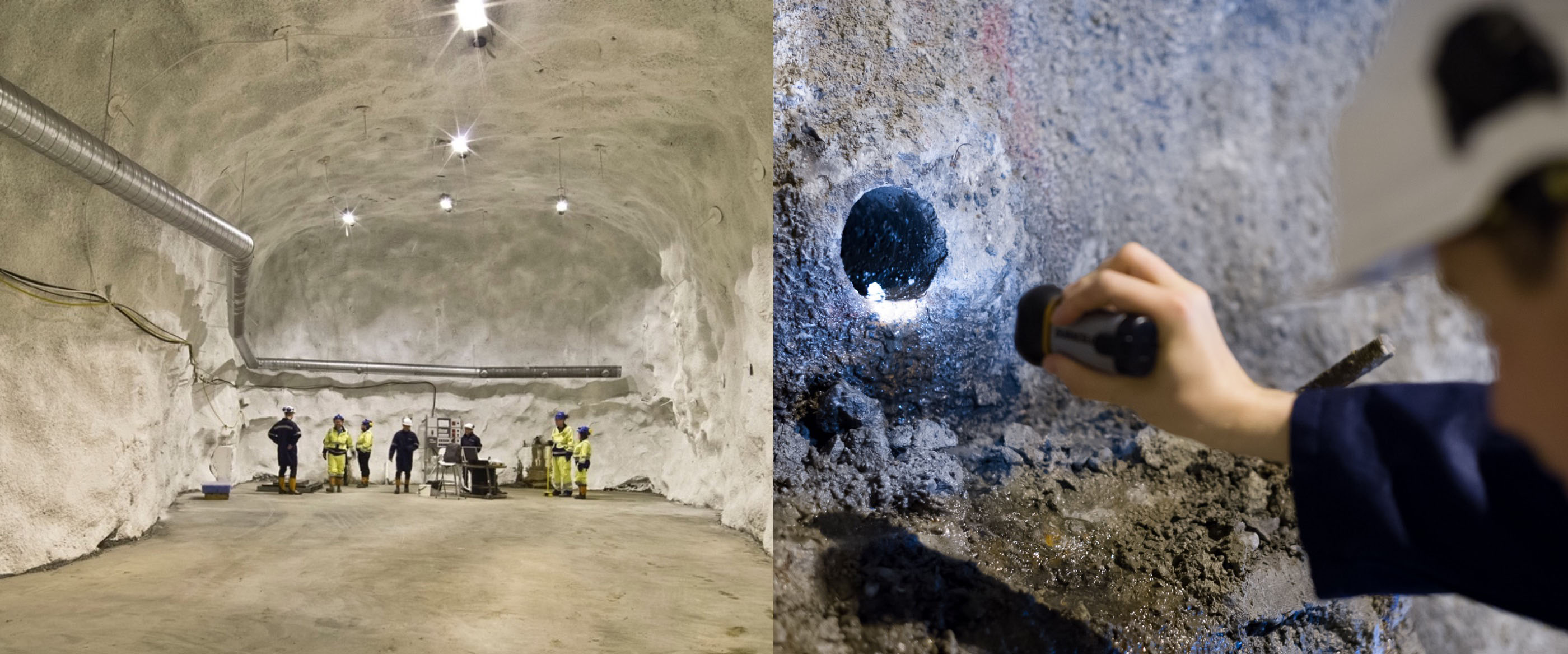}
\caption{Left: the experimental hall of Callio Lab2 after its completion in the spring of 2016. The area is about 120 m\textsuperscript{2}. The maximum height is 9 m. Right: example of a borehole originating from Callio Lab2. The diameter is 7.5 cm. Typical length is about 700 m.  Due to the high quality of the rock, the inner surface of the hole is smooth, facilitating deployment of instrumentation. The photo was taken before the walls of the main hall were painted.}
\figlab{fig-labphotos}
\end{figure}

\section{Acoustic detection of neutrinos in Callio Lab2}
\label{detection}
The borehole network, shown in \figref{fig-drillholes}, is very well suited for R\&D on the acoustic detection of neutrinos. Nearly all of the holes originate from the vicinity of Callio Lab2. The holes were produced with diamond drilling technique to extract core samples for further geological and rock mechanical studies. The drill cores have been stored and are available for further investigation, if needed. The wire-line tubes used for drilling have 75.7 mm outside diameter and consequently, 7.5 cm is the approximate diameter of the boreholes (\figref{fig-labphotos}); a typical length is about 700 m.  Due to the high quality of the rock, the inner surface of the hole is smooth, facilitating deployment of instrumentation.

The weight of the overburden exhorts significant pressure on the underlying geological formations compressing the cracks in the rock and reducing the seepage of the ground water. In the Pyh\"asalmi mine the depth separating the wet upper part from the dry lower part is at around 600 m. Consequently, the boreholes at the depth around 1400 m are classified as geologically dry. The largest observed water flow is in liters per hour.

To verify the feasibility of the proposed approach one could insert into the water-filled boreholes several hydrophones of the type used or planned to be used by ANTARES/AMADEUS \cite{9_Aguilar:2010ac} and KM3NeT \cite{10_Adrian-Martinez:2016fdl}. As a next step one could verify coupling of acoustic piezo sensors, for instance of the type described in \cite{11_Buis:2017tnx}, directly to the rock.

\section{Signal simulations}
\label{simulations}
Currently we are at the early stages of signal generation modelling and propagation in rock. Consequently, the results shown in \figref{fig-BIP} and \figref{fig-edepsim} are only indicative. To our knowledge there are no readily-available software packages capable of generating particle showers from interaction of EeV neutrinos in rock. Therefore, to obtain a reliable comparison of differences between water and rock, we have made the calculations at a lower energy. For our Geant4 \cite{12_Geant4} simulations we have used 100~TeV protons as the leading particles initiating showers in both media. Another complex process is the propagation of pressure waves in rock. Unlike water, ice or salt, rock consists of several components with multi-crystalline structure and different sound transmitting parameters. At this stage we have decided to show the pressure signal (\figref{fig-BIP}) in arbitrary units only, representing the situation at the distance of 1 km perpendicular to the shower axis without attenuation. Since the attenuation length is larger in rock than in water, the signal amplitude ratio between rock and water is expected to increase further with the distance. 

The main observation from \figref{fig-BIP} is that the expected amplitude of BIP in rock is by an order of magnitude larger than in water! The same conclusion can be derived from a frequently used expression for pressure deviation described in \cite{2_Lahmann:2016qmc}. The parameters used for comparisons are listed in \tabref{tab-params}.

\begin{table}
\centering
\caption{List of parameters used in our estimations.}
\tablab{tab-params}
\begin{tabular}{lll}
\hline
Parameter                                                                            &   Water                  &   Rock  \\\hline
Thermal expansion coefficient, $\alpha$ $\left[\frac{1}{\mathrm{K}}\right]$          &   $2.0 \times 10^{-4}$   &   $2.4 \times 10^{-5}$ \\
Specific heat, $C_p$ $\left[\frac{\mathrm{J}}{\mathrm{kg} \cdot \mathrm{K}}\right]$  &   $3.8 \times 10^{3}$    &   $0.73 \times 10^{3}$ \\
Speed of sound $\left[ \frac{\mathrm{m}}{\mathrm{s}}\right]$                         &   1500                   &   5000 \\
\hline
\end{tabular}
\end{table}

The signals also differ in the pulse length. The relevant parameter, influencing the choice of acoustic sensor, is the dominant frequency determined by the time it takes for BIP to fall from the maximum to the minimum pressure value (\figref{fig-BIP}). Our simplified model predicts about \SI{60}{\micro\second} for water and \SI{17}{\micro\second} for rock. We know of no previous simulations for rock, but our value is similar to the previously published results for Antarctic ice \cite{13_Bevan:2009kd}. Our value for water is somewhat larger than that in \cite{13_Bevan:2009kd} but this is not surprising considering the simplified assumptions used at this stage of our simulations.

\section{Main challenges}
\label{challenges}
Despite many advantages, the choice of bedrock for the acoustic detection of EeV neutrinos is far from obvious. One concern is nonuniformity of the rock and the effect of cracks and layer boundaries on the signal propagation and attenuation. The possible fault lines will certainly degrade the acoustic properties. On the other hand, the compression force of the rock overburden above will have a compensating effect by sealing the minor cracks and irregularities. For salt and sand, the improvement is significant with just a few MPa of pressure applied \cite{15_Bonner}. At the depth of 2 km the rock pressure is higher by three orders of magnitude. 

Another major concern is the acoustic background. It is clear that the microphones will pick~up a~lot~of sound signals from the rock. Fortunately, the expected neutrino signals should have a characteristic directional and frequency pattern. With the aid of suitable numerical algorithms, it should be possible to distinguish them from the background. 

Among many technical challenges are, for instance, the coupling of the sensors to the rock, optimizing the frequency response of the microphones, considering the effects of water in the boreholes, optimizing the simulations, etc.

\section{Summary and outlook}
\label{summary}
Preliminary simulations have confirmed the validity of our novel proposal to utilize bedrock as a medium for acoustic detection of particle showers following interactions of ultra-high energy neutrinos. Since the density of rock is three-times larger and the speed of sound is four-times larger compared to water, the amplitude of the generated bipolar pressure pulse in rock should be larger by an order of magnitude. Higher density also translates into the increased target mass for the neutrino interaction and hence into the increase in the event rate. In addition, rock has a noticeably longer attenuation length for sound waves than water reducing the required grid density for sensor deployment. The Pyh\"asalmi mine has a unique infrastructure and rock conditions to test this idea and, if successful, extend it to a full-size experiment. 

As part of preparations for the first round of proof-of-principle measurements, our team continues simulations and signal transmission modeling. We welcome new collaborators to participate in this research.

\bibliography{arena-2018}

\end{document}